\documentclass[%
 reprint,
 amsmath,amssymb,
prb,
]{revtex4-1}

\usepackage{dcolumn}
\usepackage{bm}

\usepackage[pdftex]{graphicx}
\usepackage{epstopdf} 
\usepackage{hyperref}
\usepackage{siunitx}
\usepackage{bm}
\usepackage{verbatim}
\usepackage[version=4]{mhchem}
\usepackage{enumitem}
\usepackage{wasysym}

\usepackage{bigints}
\usepackage{color}

\begin{document}

\preprint{AIP/123-QED}

\title{Design and characterization of a novel toroidal split-ring resonator}

\author{J.S. Bobowski}
 \email{Jake.Bobowski@ubc.ca}
\author{Hiroko Nakahara}%
\affiliation{ Department of Physics, University of British Columbia Okanagan, Kelowna, British Columbia V1V 1V7, Canada}


\date{\today}

\begin{abstract}
The design and characterization of a novel toroidal split-ring resonator (SRR) is described in detail.  In conventional cylindrical SRRs, there is a large magnetic flux within the bore of the resonator.  However, there also exists a non-negligible magnetic flux in the free space surrounding the resonator.  The energy losses associated with this radiated power diminish the resonator's quality factor.  In the toroidal SRR, on the other hand, the magnetic field lines are strongly confined within the bore of the resonator resulting in high intrinsic quality factors and stable resonance frequencies without requiring additional electromagnetic shielding.  This paper describes the design and construction of a toroidal SRR as well as an experimental investigation of its cw response in the frequency-domain and its time-domain response to an rf pulse. Additionally, the dependence of the toroidal SRR's resonant frequency and quality factor on the strength of inductive coupling to external circuits is investigated both theoretically and experimentally.
%
\end{abstract}

\maketitle

\section{\label{sec:intro}Introduction}

Split-ring resonators (SRRs) are used in a number of areas of modern experimental\cite{Burresi:2009} and applied\cite{Xiao:2007, Ricci:2006} physics.  Most notably, SRRs are critical components of two-dimensional metamaterials engineered to have simultaneously negative permittivity and permeability at microwave frequencies.\cite{Smith:2000, Shelby:2001}  Additionally, SRRs are used as devices for making high-resolution measurements of the electromagnetic (EM) properties of materials at frequencies between 10 and 2000~MHz.  Examples of the types EM measurements that have been made using SRRs include the surface resistance and magnetic penetration depth of superconducting single crystals,\cite{Bonn:1991, Hardy:1993, Bobowski:2010} the complex permittivity of dielectric materials, and the conductivity of aqueous solutions.\cite{Bobowski:2013, Bobowski:2015}  One could also use SRRs to make measurements of the complex permeability of, for example, suspensions or composite materials containing magnetic nanoparticles.\cite{Xi:2013, Zhu:2013, Bobowski:2015}  This paper describes a novel toroidal SRR designed for measurements of EM material properties.  

The paper is organized as follows: Sec.~\ref{sec:cylindrical} briefly reviews the standard cylindrical SRR.  This section also estimates the radiation resistance associated with an unshielded cylindrical SRR.  Section~\ref{sec:toroidal} introduces the toroidal SRR geometry and gives the intrinsic capacitance, inductance, and resistance of the resonator.  An experimental characterization of the resonance frequency and quality factor of a copper toroidal SRR is presented in Sec.~\ref{sec:characterize}. Both the frequency-response of the resonator and its transient response to an rf pulse are investigated.  In Sec.~\ref{sec:compare}, experimental measurements are used to directly compare and contrast cylindrical and toroidal SRRs.  Section~\ref{sec:coupling} explores, both theoretically and experimentally, the effect that inductive coupling has on the SRR's resonance frequency and quality factor. Section~\ref{sec:summary} provides a summary and discusses future applications of the toroidal SRR.

\section{Cylindrical Split-Ring Resonators}\label{sec:cylindrical}
\begin{figure*}
\begin{tabular}{lclc}
(a) & \includegraphics[keepaspectratio, width=.92\columnwidth]{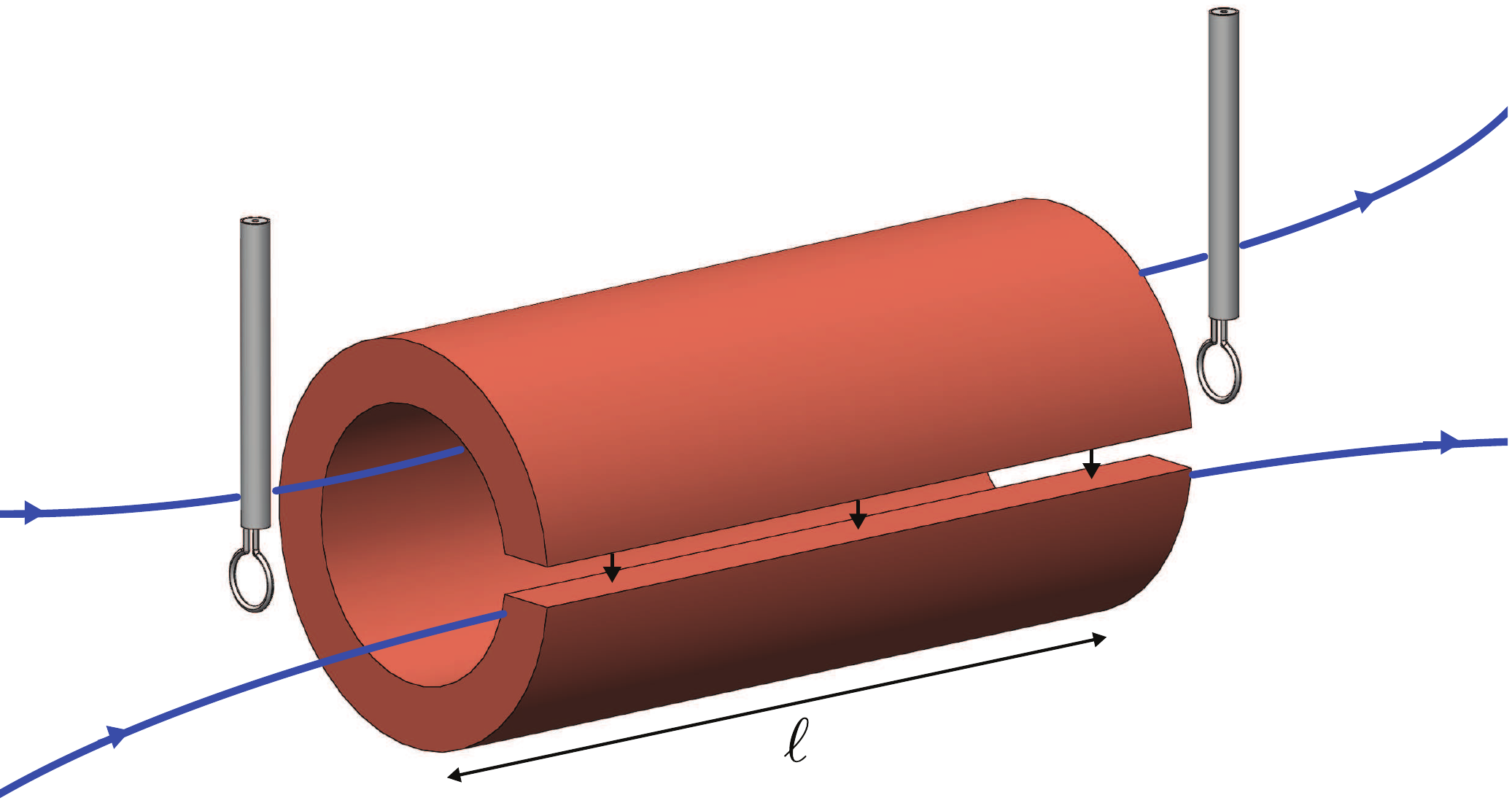} & (c) & \includegraphics[keepaspectratio, width=.92\columnwidth]{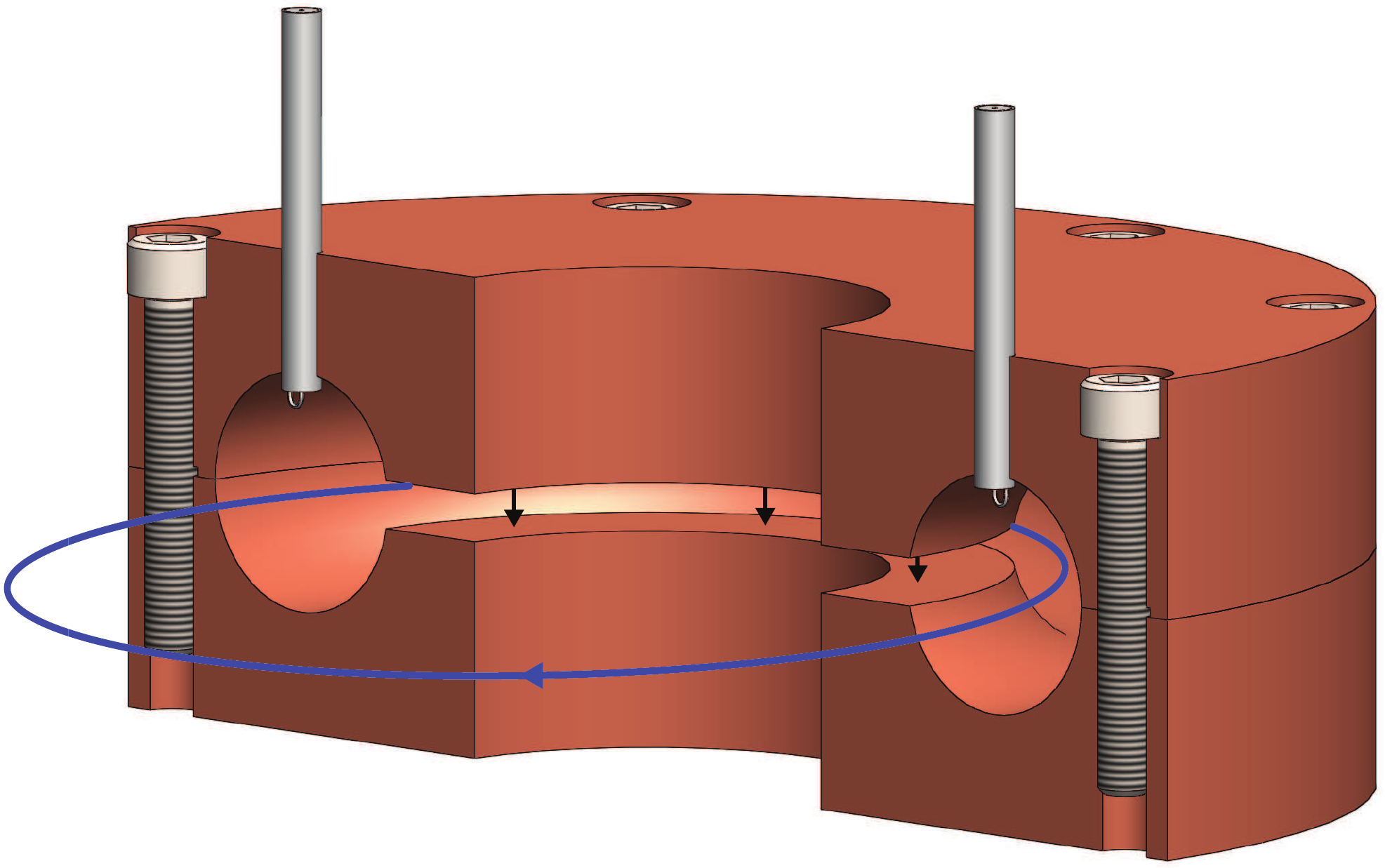}\\
(b) & \includegraphics[keepaspectratio, width=0.35\columnwidth]{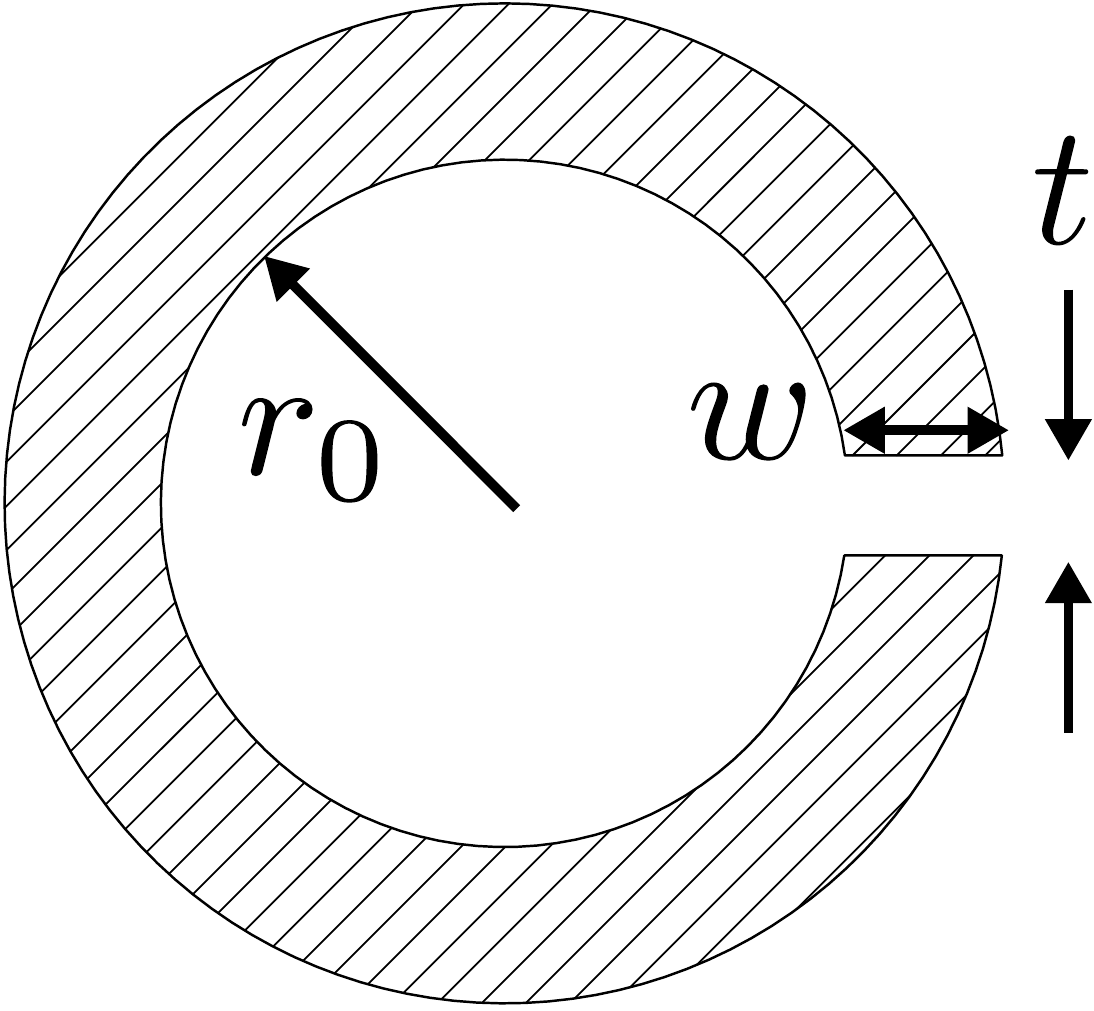} & (d) & \includegraphics[clip=true, keepaspectratio, width=0.92\columnwidth]{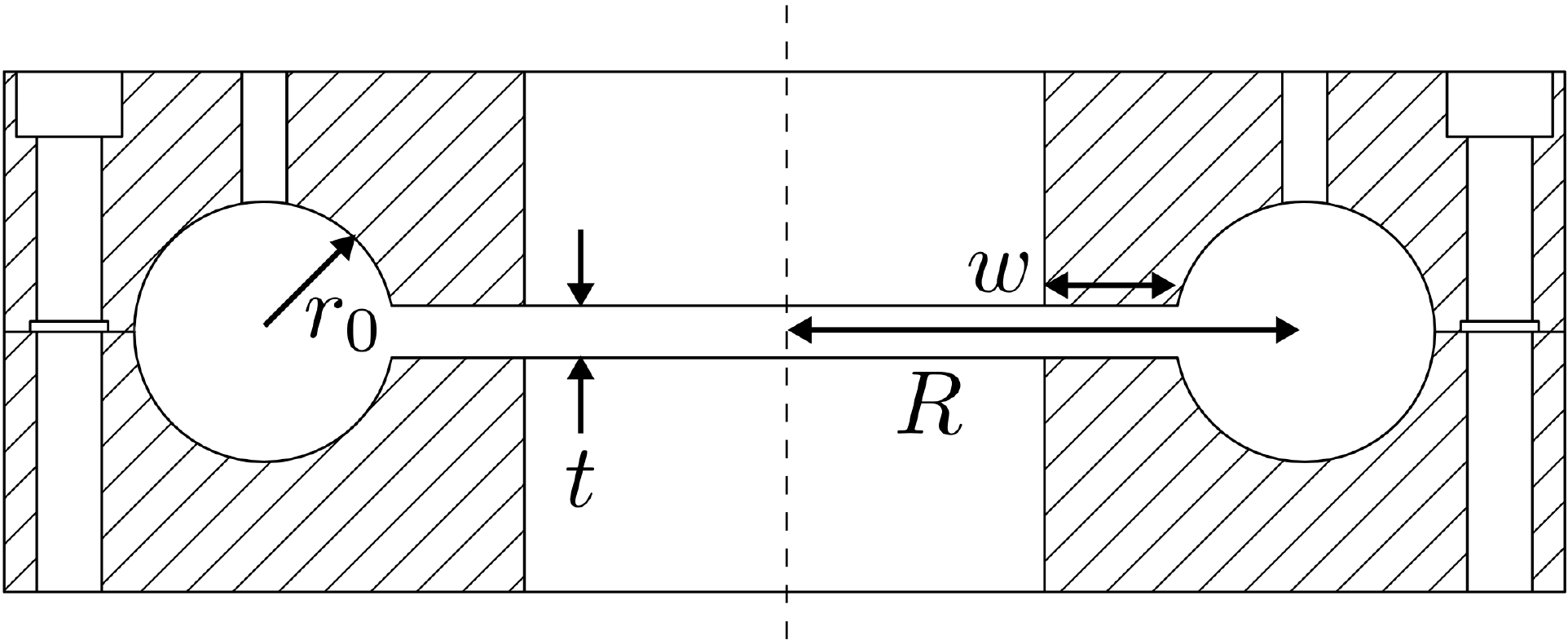}
\end{tabular}
\caption{\label{fig:SRRgeo}SRR geometries and representative electric field lines (black) in the gap and magnetic field lines (blue) in the bore of the resonators.  (a) A cylindrical SRR and two inductive coupling loops.  Magnetic field lines extend outside the bore of the cylindrical SRR.  (b) A cross-sectional view of the cylindrical SRR with the critical dimensions labelled. (c) The toroidal SRR geometry.  For clarity, half of the SRR has been cut away in this view. The coupling loops pass through a pair of holes in the top of the resonator.  In this geometry, there is very little magnetic flux outside the bore of the resonator. (d) A drawing of the cross-section of the toroidal SRR that was built with the critical dimensions labelled.  Other than the gap dimensions $t$, which have been exaggerated by a factor of ten for clarity, the drawings are to scale. To set the scales, we note that the $r_0$ values of the cylindrical and toroidal SRRs investigated in this work are 11/16~in.\ (1.75~cm) and 0.250~in.\ (0.635~cm), respectively.}
\end{figure*}

A cylindrical SRR is made by cutting a slit along the length of a conducting tube. See Fig.~\ref{fig:SRRgeo}(a).  Using the dimensions labelled in Fig.~\ref{fig:SRRgeo}(b), the approximate effective capacitance and inductance of a cylindrical SRR suspended in air are \mbox{$C_\mathrm{c}\approx \varepsilon_0 w\ell/t$} and \mbox{$L_\mathrm{c}\approx \mu_0\pi r_0^2/\ell$}. As a result, the SRR acts as an $LC$-resonator with a resonant frequency given by\cite{Hardy:1981,Bobowski:2013}
\begin{equation}
f_{0,\mathrm{c}}\approx\frac{1}{2\pi}\frac{1}{\sqrt{L_\mathrm{c}C_\mathrm{c}}}=\frac{c}{2\pi r_0}\sqrt{\frac{t}{\pi w}}\label{eq:f0c}
\end{equation}
where $c=\left(\varepsilon_0\mu_0\right)^{-1/2}$ is the vacuum speed of light.

As shown in Fig.~\ref{fig:SRRgeo}(a), a pair of inductive coupling loops are placed at either end of the SRR.  The coupling loops, made by shorting the centre conductor of a semi-rigid coaxial cable to its outer conductor, are used to couple magnetic flux into and out of the bore of the resonator.  When an rf signal is applied to the drive (input) coupling loop, circulating currents are induced on the inner surface of the SRR.  The effective resistance of the cylindrical SRR is given by\cite{Bobowski:2013}
\begin{equation}
R_\mathrm{c}\approx \rho\frac{2\pi r_0}{\delta \ell}\label{eq:Rc}
\end{equation} 
where $\rho$ is the resistivity of the conductor used to make the SRR, $\delta=\sqrt{2\rho/\left(\mu_0\omega\right)}$ is the electromagnetic skin depth, and $\omega=2\pi f$.  The resulting quality factor of the cylindrical SRR is
\begin{equation}
Q_\mathrm{c}\approx\frac{1}{R_\mathrm{c}}\sqrt{\frac{L_\mathrm{c}}{C_\mathrm{c}}}=\frac{r_0}{\delta}.\label{eq:Qc}
\end{equation}  

Quality factors close to the values predicted by Eq.~\ref{eq:Qc} are only obtained if the cylindrical SRR is surrounded by an EM shield.\cite{Hardy:1981, Bobowski:2013}  Without shielding, lower-than-expected $Q$ values are obtained due to radiative power losses.  Figure~\ref{fig:SRRgeo}(a) shows that, while the magnetic flux is concentrated within the bore of the cylindrical SRR, magnetic field lines also radiate out from the resonator and into free space.  This effect can be modelled by including a ``radiation resistance'' $R_\mathrm{r}$ in series with the previously calculated $R_\mathrm{c}$.  An order-of-magnitude estimate of the radiation resistance can be obtained by considering the power radiated by a current loop.\cite{Griffiths:1999}  In the limit that the wavelength of the radiation is   much larger than the loop radius, the radiation resistance can be estimated using\cite{Snoke:2015}
\begin{equation}
R_\mathrm{r}\approx\frac{\pi}{6}\sqrt{\frac{\mu_0}{\varepsilon_0}}\left(\frac{\omega r_0}{c}\right)^4.
\end{equation}

As an example, the estimated radiation resistance of the aluminum cylindrical SRR used both in this work and in Ref.~\onlinecite{Bobowski:2013} is eight times larger than the value of $R_\mathrm{c}$ calculated using Eq.~\ref{eq:Rc}.  To operate a cylindrical SRR at, or near, the highest achievable $Q$ values, additional EM shielding is required to suppress radiation effects.

\section{Toroidal Split-Ring Resonators}\label{sec:toroidal}

Cross-sectional views of a novel toroidal SRR geometry are shown in Figs.~\ref{fig:SRRgeo}(c) and (d).  One can imagine forming the toroidal SRR by bending the cylindrical SRR into a circle such that its two ends meet.  The obvious advantage of the toroidal geometry is that the magnetic flux is strongly confined within the bore of the resonator.  As a result, radiative effects are expected to be negligible such that high-$Q$ and high-stability resonators can be built without requiring additional EM shielding.

\subsection{Design}\label{sec:design}

\begin{figure*}
\begin{tabular}{lclc}
(a) & \includegraphics[keepaspectratio, width=.97\columnwidth]{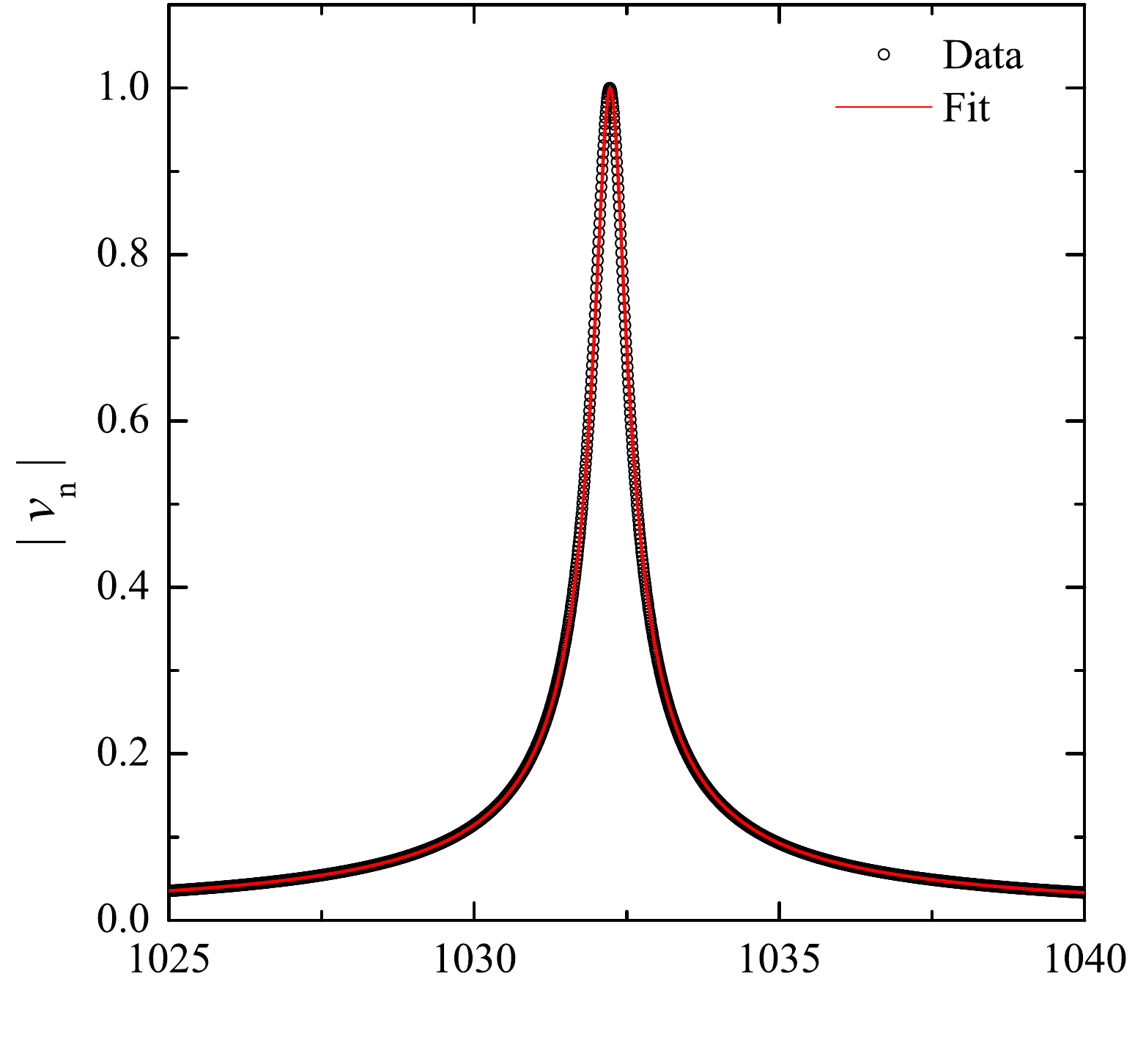} & (b) & \includegraphics[keepaspectratio, width=.95\columnwidth]{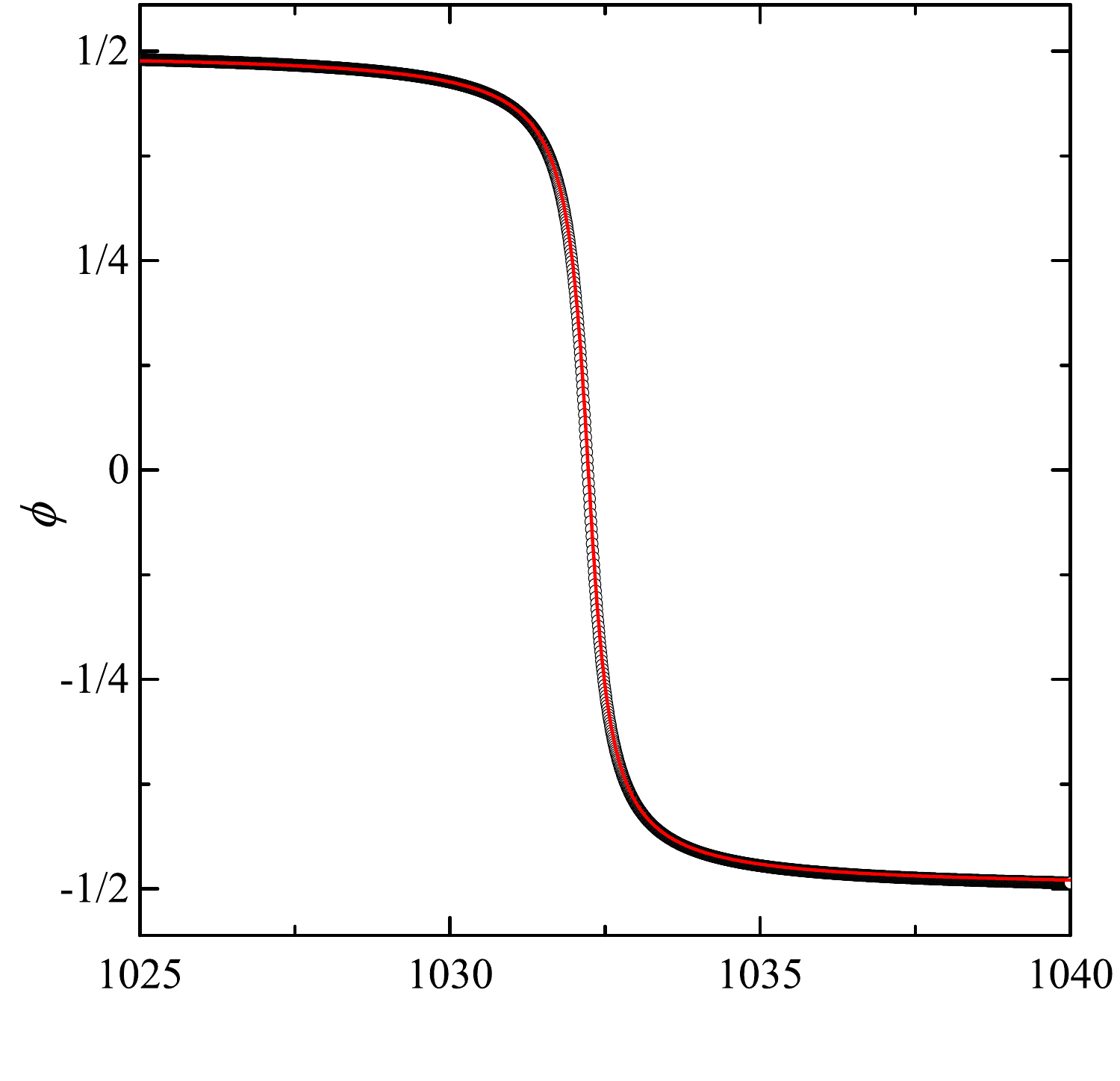}
\end{tabular}
\caption{\label{fig:vna}The (a) magnitude and (b) phase of the signal coupled out of the toroidal SRR as a function of frequency.}
\end{figure*}

Our copper toroidal SRR was made from two mating pieces.  A semicircular groove was cut into a pair of copper rings using a ball-nose end mill.  The groove dimensions were $r_0=0.250$~in.\ and $R=1.00$~in.\ (2.54~cm).  The copper rings themselves were 0.500~in.\ (1.27~cm) tall and had outer and inner diameters of 3.00~in.\ (7.62~cm) and 1.00~in.\ respectively such that \mbox{$w=0.250$}~in.  For radii less than $R-r_0$, the ring heights were reduced such that a gap forms along the inner diameter of the resonator when the two parts mate.   The size of the gap formed was \mbox{$t=0.010$}~in.\ (0.254~mm).  

A series of eight \#4-40 stainless steel bolts along the outer diameter of the resonator were used to securely mate the two parts.  Counterbored \#4 clearance holes were drilled into the top ring and \#4-40 holes drilled and tapped into the bottom ring.  As shown in Fig.~\ref{fig:SRRgeo}(d), a 0.150~in.\ (0.381~cm) wide and 0.020~in.\ (0.508~mm) deep relief was cut along the outer diameter of the top ring.  This groove restricts contact between the two mating parts to a pair of 0.050~in.\ (1.27~mm) wide feet and is designed to promote good electrical and mechanical contact between the two parts. Two holes drilled through opposite sides the top half of the toroidal SRR accommodate coupling loops made using 0.085~in.\ (2.16~mm) diameter semi-rigid coaxial cable.

\subsection{Electromagnetics}\label{sec:EM}

The capacitance $C_\mathrm{t}$ of the toroidal SRR is given by
\begin{eqnarray}
C_\mathrm{t}&=&\frac{\varepsilon_0\pi}{t}\left[\left(R-r_0\right)^2-\left(R-r_0-w\right)^2\right]\nonumber\\
&=& \frac{\varepsilon_0 \pi R^2}{t}\left(1-\frac{r_0}{R}\right)^2\left[1-\left(1-\frac{w}{R-r_0}\right)^2\right].\label{eq:Ct}
\end{eqnarray} 
In the limit that $R\gg r_0, w$; this result reduces to that of a cylindrical SRR of length $\ell=2\pi R$.

The magnetic flux $d\Phi$ through a strip of width $dr$ at a position \mbox{$R-r_0<r<R+r_0$} within the bore of the resonator is given by
\begin{equation}
d\Phi=\left(\frac{\mu_0 i}{2 \pi r}\right)2 r_0\sqrt{1-\left(\frac{R-r}{r_0}\right)^2}\, dr
\end{equation}
where $i$ is the SRR current.  The total inductance is determined from $\Phi/i$ such that
\begin{eqnarray}
L_\mathrm{t}&=&\frac{\mu_0 r_0^2}{\pi R}\bigintsss_{-1}^{1}\frac{\sqrt{1-z^2}}{1-\dfrac{r_0}{R}z}\, dz\nonumber\\
&=&\mu_0 R\left[1-\sqrt{1-\left(\frac{r_0}{R}\right)^2}\right]\label{eq:Lt}
\end{eqnarray}
where $z\equiv (R-r)/r_0$.  Once again, in the limit $R\gg r_0$, this result reduces that of a cylindrical SRR of length $\ell=2\pi R$.

The effective resistance $R_\mathrm{t}$ of the toroidal SRR is calculated in the Appendix and is given by Eq.~\ref{eq:A}.  The result, repeated here, is given in terms of the following elliptical integral
\begin{equation}
R_\mathrm{t}=\frac{\rho \, r_0}{2\pi R\,\delta}\bigintsss_0^{2\pi}\frac{d\theta}{\sqrt{1+\left(r_0/R\right)^2-2\left(r_0/R\right)\cos\theta}}.\label{eq:Rt}
\end{equation}
In the large-$R$ limit, the integral evaluates to $2\pi$ such that the cylindrical SRR result with $\ell=2\pi R$ is once again recovered.

Using Eqs.~\ref{eq:Ct} to \ref{eq:Rt}, the dimensions given in Sec.~\ref{sec:design}, and assuming a copper resistivity of \mbox{$\rho=1.68~\mu\Omega\,\mathrm{cm}$}; the expected resonant frequency and $Q$ of the toroidal SRR studied in this work are 1.06~GHz and 3200, respectively.

\section{Frequency \& Transient Response}\label{sec:characterize}

\subsection{Frequency Response}\label{sec:freq}

The frequency response of the toroidal SRR was characterized using an Agilent N5241A Vector Network Analyzer (VNA).  The drive coupling loop was connected to the VNA output port and used to excite the SRR.  On resonance, relatively large currents are induced on the surface of the resonator bore.  This current, in turn, generates a magnetic flux that is detected by the receiver (output) coupling loop connected to the VNA input port.  The strength of the detected signal is directly proportional to the magnitude of the current, and therefore, inversely proportional to effective impedance of the SRR.

As an $LRC$ resonator, the SRR is expected to have a Lorentzian frequency response such that
\begin{eqnarray}
\left|v_\mathrm{n}\right|&=&\frac{1}{\sqrt{1+Q^2\left(\dfrac{f_0}{f}-\dfrac{f}{f_0}\right)^2}}\label{eq:mag}\\
\tan\phi &=& Q\left(\frac{f_0}{f}-\frac{f}{f_0}\right)\label{eq:phi}
\end{eqnarray}
where $\left| v_\mathrm{n}\right|$ represents a normalized signal amplitude and $\phi$ is the phase difference between the input and output signals.  The measured magnitude and phase of the toroidal SRR signal are shown in Fig.~\ref{fig:vna}.  

When analyzing the VNA data, the lengths of coaxial cable used to make the coupling loops were treated as ideal transmission lines.  Because the coaxial cables were assumed to be lossless, the magnitude data in Fig.~\ref{fig:vna}(a) are raw data obtained directly from the VNA.  However, the $\phi$ data in Fig.~\ref{fig:vna}(b) were corrected for the phase shift that occurs due to the length of the cables used to construct the coupling loops.  The data were corrected by removing the \mbox{$\Delta \phi=\sqrt{\varepsilon_\mathrm{r}}\omega\ell/c$} contribution arising from cables of length $\ell$ and dielectric constant $\varepsilon_\mathrm{r}$.  The total length of the cable used to make the pair of coupling loops was approximately 20~cm.

The magnitude and phase data were simultaneously fit the Lorentzian $\left\vert v_\mathrm{n}\right\vert$ and $\phi$ expressions above.  The fits are remarkably good and result in parameter values \mbox{$f_0=1.032$}~GHz and \mbox{$Q=2023\pm 2$}.  The uncertainty in $f_0$ is in the range of tens of kilohertz.  The measured resonance frequency is very close to the predicted value calculated in Sec.~\ref{sec:EM}.  The measured $Q$, however, is only about two-thirds of the predicted value.  

Lower-than-expected $Q$ values are typical and can be attributed to a number of factors.  First, the resistivity of the copper used to build the resonator may be greater than the assumed value of \mbox{$\rho=1.68~\mu\Omega\,\mathrm{cm}$}.  Second, the toroidal SRR used in this study was made from two parts.  There can be an additional effective resistance associated with the mechanical and electrical connection between the two parts.  Just prior to assembling the resonator, it is good practice to polish the mating surfaces with fine emery paper or steel wool to remove surface oxide layers.  Finally, as will be discussed in Sec.~\ref{sec:coupling}, the $Q$ of the resonator also depends on how strongly the resonator is coupled to the VNA (or signal generator/analyzer) output and input.  For the toroidal SRR, the coupling strength can be tuned by raising/lowering the height of the coupling loops shown in Fig.~\ref{fig:SRRgeo}(c).  In the figure, the coupling loops are entirely within the bore of the resonator which corresponds to maximum coupling for a fixed loop size.  The data shown in Fig.~\ref{fig:vna} were obtained using maximum coupling in order to achieve the best possible signal-to-noise ratio.  Section~\ref{sec:coupling} will show that $Q$-values in excess of 2700 were obtained under weak-coupling conditions.

\subsection{Transient Response}\label{sec:trans}
The resonance frequency and quality factor of the toroidal SRR can also be characterized by studying its transient response to a resonant rf pulse.  For a highly-underdamped resonator ($Q\gg 1$), the voltage coupled out of the SRR is expected to be a damped sinusoidal signal of the form
\begin{equation}
v(t)=V_0 e^{-t/\tau}\sin\left(\omega_0 t+\alpha\right)\label{eq:ring}
\end{equation}
where $V_0$ is the amplitude at time $t=0$, $\alpha$ is a phase constant, and  $\tau=2Q/\omega_0$ is the decay time constant.

\begin{figure}[t]
\includegraphics[keepaspectratio, width=\columnwidth]{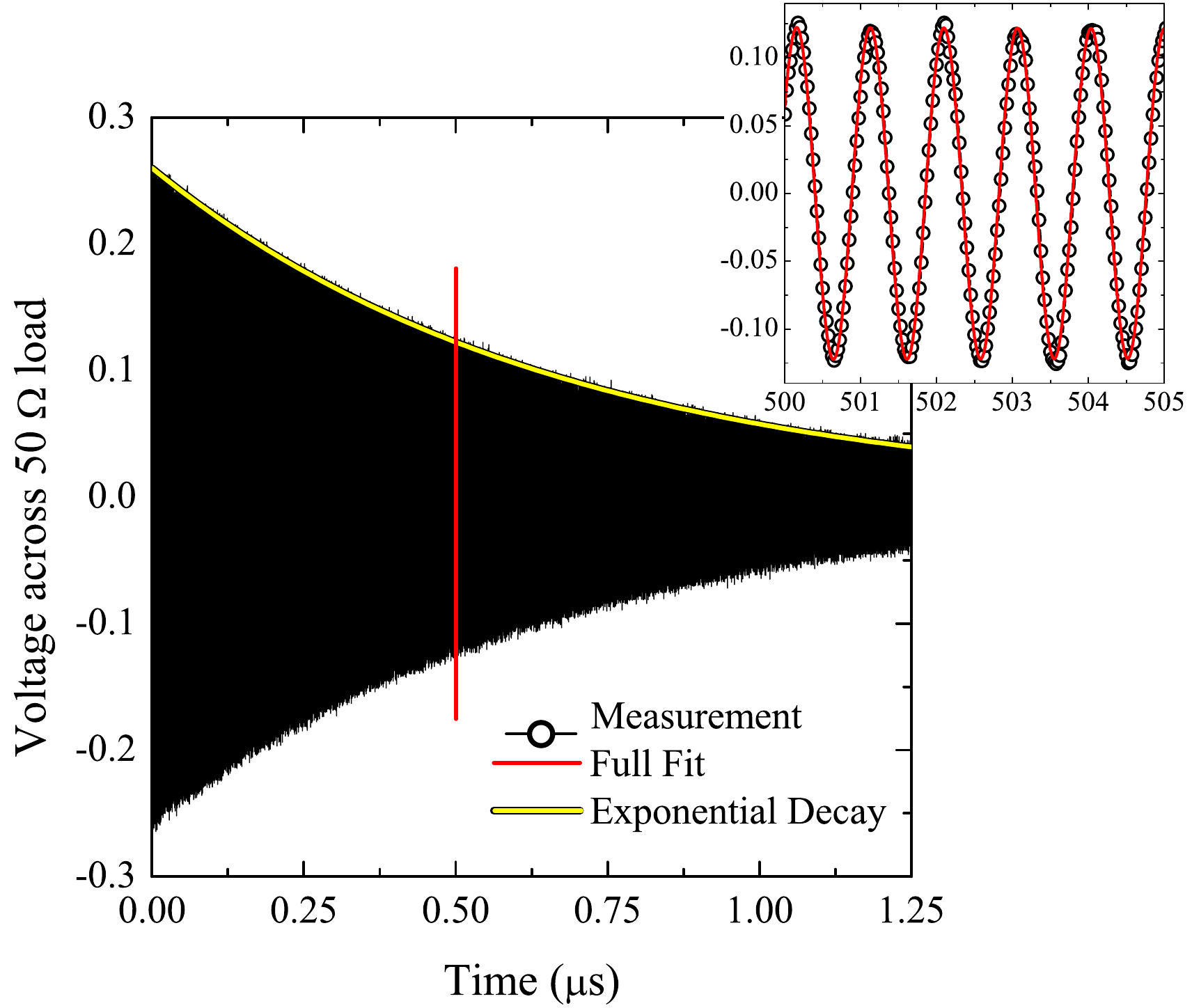}
\caption{\label{fig:ring}Transient response of the toroidal SRR signal to a resonant rf pulse applied to the input coupling loop. The rf power coupled out of the resonator decays exponentially.  The inset shows a detailed view of the measured signal (black circles) and the best-fit line (red) approximately $0.5~\mu s$ after the end of the pulse.}
\end{figure}
To measure the transient response of the toroidal SRR, a Rohde \& Schwarz SMY02 signal generator with pulse modulation was used to apply a pulse to the input coupling loop and the signal coupled out of the resonator was measured using a  Tektronix DP070804B (7~GHz) Digital Phosphor Oscilloscope.  The pulse width used (11.6~$\mu$s) was sufficiently long to ensure that the current induced in the resonator reached equilibrium before the end of the pulse.  The rf frequency of the pulse (1.032~GHz) was adjusted to the known resonance frequency of the SRR and the rf power was set to 19~dBm (79~mW).  Finally, to achieve the maximum possible signal strength, the coupling loops were positioned within the bore of the resonator (similar to the loop positions shown in Fig.~\ref{fig:SRRgeo}(c)).

The measured transient response of the toroidal SRR following the end of a pulse is shown in Fig.~\ref{fig:ring}.  The expected ``ringing'' behaviour with an exponentially decaying amplitude is clearly observed.  The inset of the figure shows several oscillations of the signal approximately 0.5~$\mu$s after the end of a pulse.   The complete dataset was fitted to Eq.~\ref{eq:ring}.  The fit was excellent and produced parameter values $f_0=1.032$~GHz (uncertainty in the kilohertz range) and $Q=2157\pm 1$.  The slightly larger value of $Q$ measured in this section (2157) compared to the previous section (2023) is an indication that better electrical contact between the two halves 
of the toroidal SRR was achieved during the second measurement.

\section{Cylindrical Versus Toroidal SRR\lowercase{s}}\label{sec:compare}

This section compares and contrasts cylindrical and toroidal SRRs.  The cylindrical SRR used was first reported on in Ref.~\onlinecite{Bobowski:2013}.  The aluminum resonator was made from two halves which bolted together such that \mbox{$r_0=11/16$}~in., \mbox{$w=5/16$}~in.\ (0.794~cm), \mbox{$t=0.010$}~in., and \mbox{$\ell=4$}~in.\ (10.2~cm).  For these measurements, the input coupling loops were driven using the Rohde \& Schwarz SMY02 signal generator and the signal coupled out of the resonators were detected using an Anritsu MS610C2 spectrum analyzer.  Data collection was automated using a LabVIEW program written in-house.

Due to their different geometries, the two SRRs have quite different resonant frequencies: 340~MHz and 1032~MHz for the shielded cylindrical SRR and the toroidal SRR, respectively.  In order to make a fair comparison of these two resonators, the detected signals were plotted as a function of $f/f_0-1$.  In this way, the shapes of the data obtained were determined solely by the quality factors of the resonators.

\begin{figure}[t]
\includegraphics[keepaspectratio, width=0.94\columnwidth]{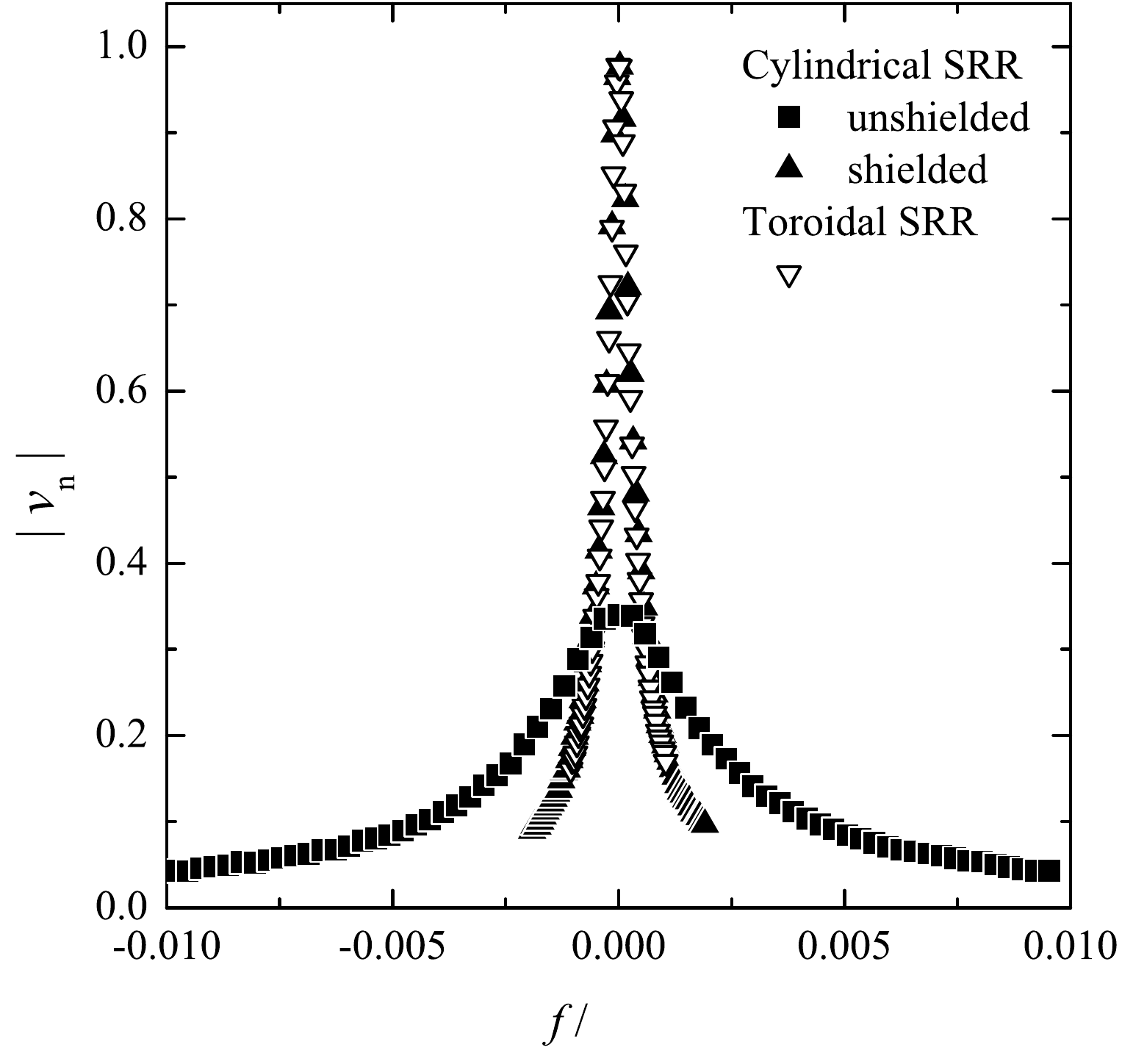}
\caption{\label{fig:cylVStor}Comparison of the toroidal SRR resonance (open triangles, \mbox{$Q=2497$}) to that of the unshielded (solid squares, \mbox{$Q=357$}) and shielded (solid triangles, \mbox{$Q=2299$}) cylindrical SRR.  The signal amplitudes of the toroidal SRR and shielded cylindrical SRR datasets were scaled such that, on resonance, the normalized signal amplitudes equal one.  For the unshielded cylindrical SRR, using the same coupling strength, an additional 9~dBm of power was required to reach the same peak signal amplitude as the shielded dataset.}
\end{figure}

Figure~\ref{fig:cylVStor} shows the resonances of the cylindrical SRR (with and without additional EM shielding) and the toroidal SRR.   
The unshielded measurements of the cylindrical SRR were made by suspending the resonator in air using a pair of Teflon rods and, as shown in Fig.~\ref{fig:SRRgeo}(a), placing coupling loops one inch from either end of the resonator.  The shielded measurements were made by suspending the SRR within a 97~cm long plastic tube that was wrapped with several layers of aluminum foil.  The outside diameter of the plastic tube was 21.4~cm.\cite{Bobowski:2013}  As shown in Fig.~\ref{fig:cylVStor}, the $Q$ of the shielded resonator is more than six times greater than that of the unshielded resonator.  The EM shield prevents magnetic flux from emanating out into free space thereby suppressing the radiation resistance discussed in Sec.~\ref{sec:cylindrical}.  On the other hand, magnetic field lines are strongly confined with the bore of the toroidal SRR such that high quality factors are obtained without requiring additional shielding.

We also note that the cylindrical SRR's resonance frequency can be shifted substantially by the presence of nearby dielectrics, ferromagnetic materials, or conductors.  These objects interfere with the EM fields surrounding the cylindrical resonator and modify the effective capacitance and/or inductance of the device.  In contrast, the only substantial EM fields outside the bore and gap of the toroidal SRR exist within the inner diameter \mbox{$(0<r<R-r_0-w)$} of the pair mating of copper rings.  It is easy to exclude objects from this small and isolated region of space and therefore achieve very stable resonance frequencies.  The toroidal SRR geometry introduced in this work has inherently high quality factors and stable resonance frequencies.  Cylindrical SRRs can attain similar attributes only if they are surrounded by bulky EM shields.

Compact, high-$Q$, and high-stability toroidal SRRs can certainly be used for accurate and precise measurements of the EM properties of materials.  However, these resonators could also find use in metamaterial research.  Thus far, excessive losses have prevented researches from developing many of the applications anticipated from engineered metamaterials.\cite{Boardman:2011}  A number of techniques to compensate for losses by introducing gain mechanisms into these system are currently being investigated.\cite{Boardman:2011, Lapine:2014}  Another area of current metameterial research is in the development of tunable ``meta-atoms''.  For example, the resonance frequency of SRRs loaded with varactor diodes can be tuned by controlling the biasing of the diode.\cite{Kapitanova:2012, Slobozhanyuk:2014}  Meta-atoms based on the toroidal SRR may be possible.  The inherently high quality factor of these resonators would be a clear advantage.  However, there are also challenges.  First, the toroidal geometry is much bulkier than the current SRRs that are used in metamaterials that operate at microwave frequencies.  Second, a method to introduce coupling between adjacent toroidal SRRs would have to be developed.  The current toroidal SRR design does not allow electric or magnetic flux from one resonator to couple to a second.  However, we note that researchers have recently demonstrated a novel optical coupling between a pair of SRRs that could also be applied to toroidal SRRs.  In their work, coupling was achieved using an LED and a photodiode while direct electromagnet coupling was suppressed by cross-polarizing the pair of SRRs.\cite{Slobozhanyuk:2014}

\section{Effect of Inductive Coupling}\label{sec:coupling}

To this point, the SRR has been modelled as a series-$LRC$ circuit without taking into account the effects of the inductive coupling.  The first part of this section develops and analyzes a more complete circuit model that includes coupling effects.  The results obtained can be applied to either the cylindrical or toroidal SRR geometries.  The second part of this section experimentally investigates how the toroidal SRR's resonant frequency and quality factor depend on the strength of the inductive coupling.  Finally, the section concludes with a brief discussion of capacitive coupling methods.

\subsection{Circuit Model}

\begin{figure}[t]
\includegraphics[keepaspectratio, width=0.85\columnwidth]{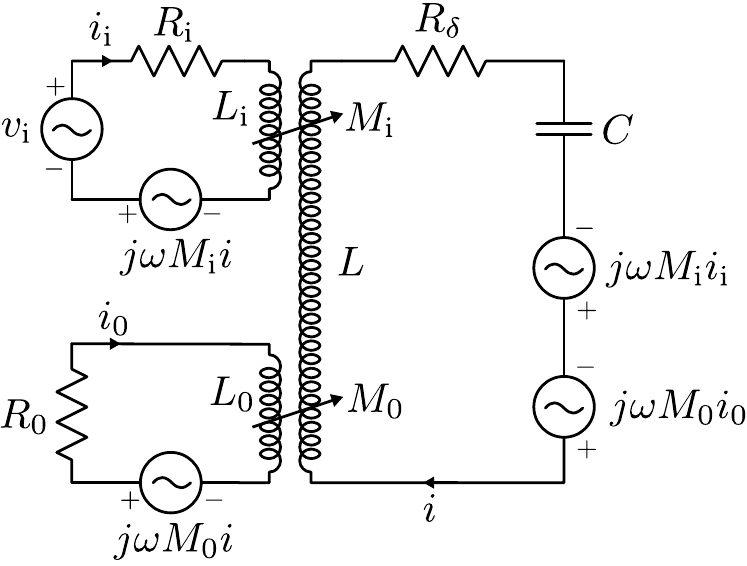}
\caption{\label{fig:couplingCircuit}Equivalent circuit of a SRR inductively coupled to both a signal generator supplying $v_\mathrm{i}$ with output impedance $R_\mathrm{i}$ and a receiver with input impedance $R_0$.}
\end{figure}

Figure~\ref{fig:couplingCircuit} shows the equivalent circuit that is used to model the SRR with inductive coupling to both a signal generator and a signal analyzer included.\cite{Momo:1983}
In the figure $L$, $C$, and $R_\delta$ are the SRR inductance, capacitance, and resistance.  The expressions for the cylindrical and toroidal geometries are given in Secs.~\ref{sec:cylindrical} and \ref{sec:EM}, respectively.  $L_\mathrm{i}$ is the inductance of the input/drive coupling loop connected to a signal generator with output impedance $R_\mathrm{i}$ and supplying signal $v_\mathrm{i}$.  Likewise, $L_0$ is the inductance of the output/receiver coupling loop connected to a signal analyzer with input impedance $R_0$.  In rf applications, one typically has \mbox{$R_\mathrm{i}=R_0=50~\Omega$}.  The mutual inductance between the input and output coupling loops and the SRR inductance are denoted $M_\mathrm{i}$ and $M_0$, respectively.

Due to mutual inductance $M_\mathrm{i}$, current $i_\mathrm{i}$ results in an induced emf $j\omega M_\mathrm{i}i_\mathrm{i}$ in the SRR circuit and current $i$ generates an induced emf $j\omega M_\mathrm{i} i$ in the signal generator circuit.  Likewise, mutual inductance $M_0$ results in an induced emf in the signal receiver circuit and an additional emf in the SRR circuit.  In Fig.~\ref{fig:couplingCircuit}, the current directions and induced emf polarities have been drawn assuming that $v_\mathrm{i}$ is instantaneously increasing with the polarity indicated in the figure.

A Kirchhoff analysis of the three loops of the circuit leads to the following system of three equations
\begin{eqnarray}
v_\mathrm{i}+j\omega M_\mathrm{i} i &=& i_\mathrm{i} \left(R_\mathrm{i}+j\omega L_\mathrm{i}\right)\label{eq:1}\\
j\omega M_0 i &=& i_0 \left(R_0+j\omega L_0\right)\label{eq:2}\\
j\omega\left(M_\mathrm{i} i_\mathrm{i}+M_0 i_0\right) &=& i\left[R_\delta +j\omega L+\frac{1}{j\omega C}\right].\label{eq:3}
\end{eqnarray} 
Using Eqs.~\ref{eq:1} and \ref{eq:2} to eliminate $i_\mathrm{i}$ and $i_0$ in Eq.~\ref{eq:3} allows one to write an expression for $i$ that is of the form
\begin{eqnarray}
v_\mathrm{eff}=i\left[R_\mathrm{eff}+j\omega L_\mathrm{eff}+\frac{1}{j\omega C}\right]
\end{eqnarray}
where
\begin{eqnarray}
\frac{v_\mathrm{eff}}{v_\mathrm{i}}&=&\frac{\omega^2 L_\mathrm{i} M_\mathrm{i}/R_\mathrm{i}^2}{1+\left(\omega L_\mathrm{i}/R_\mathrm{i}\right)^2}\left(1-\frac{R_\mathrm{i}}{j\omega L_\mathrm{i}}\right)\\
\frac{R_\mathrm{eff}}{R_\delta}&=& 1+\frac{\left(\omega M_\mathrm{i}/R_\mathrm{i}\right)^2}{1+\left(\omega L_\mathrm{i}/R_\mathrm{i}\right)^2}\frac{R_\mathrm{i}}{R_\delta}+\frac{\left(\omega M_0/R_0\right)^2}{1+\left(\omega L_0/R_0\right)^2}\frac{R_0}{R_\delta}\\
\frac{L_\mathrm{eff}}{L}&=& 1-\frac{\left(\omega M_\mathrm{i}/R_\mathrm{i}\right)^2}{1+\left(\omega L_\mathrm{i}/R_\mathrm{i}\right)^2}\frac{L_\mathrm{i}}{L}-\frac{\left(\omega M_0/R_0\right)^2}{1+\left(\omega L_0/R_0\right)^2}\frac{L_0}{L}.\qquad
\end{eqnarray}
For very small coupling loops, such as those shown in Fig.~\ref{fig:SRRgeo}(c), the loop inductance is expected to be very small such that \mbox{$\omega L_\mathrm{i}/R_\mathrm{i},~\omega L_0/R_0\ll 1$} and
\begin{eqnarray}
\frac{v_\mathrm{eff}}{v_\mathrm{i}}&\approx &\frac{\omega M_\mathrm{i}}{j R_\mathrm{i}}\\
\frac{R_\mathrm{eff}}{R_\delta}&\approx & 1+\left(\frac{\omega M_\mathrm{i}}{R_\mathrm{i}}\right)^2\frac{R_\mathrm{i}}{R_\delta}+\left(\frac{\omega M_0}{R_0}\right)^2\frac{R_0}{R_\delta}\\
\frac{L_\mathrm{eff}}{L}&\approx & 1-\left(\frac{\omega M_\mathrm{i}}{R_\mathrm{i}}\right)^2\frac{L_\mathrm{i}}{L}-\left(\frac{\omega M_0}{R_0}\right)^2\frac{L_0}{L}.
\end{eqnarray}
Finally, the effective resonance frequency \mbox{$\omega_\mathrm{eff}\approx 1/\sqrt{L_\mathrm{eff}C}$} and quality factor \mbox{$Q_\mathrm{eff}\approx R_\mathrm{eff}^{-1}\sqrt{L_\mathrm{eff}/C}$} can be calculated.  In the same small $L_\mathrm{i}$ and $L_0$ limits used above and, keeping only terms of order $M_\mathrm{i}^2$ and $M_0^2$, the results are
\begin{eqnarray}
\frac{\omega_\mathrm{eff}}{\omega_0}&\approx & 1+\frac{1}{2}\left(\frac{\omega M_\mathrm{i}}{R_\mathrm{i}}\right)^2\frac{L_\mathrm{i}}{L}+\frac{1}{2}\left(\frac{\omega M_0}{R_0}\right)^2\frac{L_0}{L}\\
\frac{Q_\mathrm{eff}}{Q}&\approx &1-\left(\frac{\omega M_\mathrm{i}}{R_\mathrm{i}}\right)^2\left(\frac{R_\mathrm{i}}{R_\delta}+\frac{1}{2}\frac{L_\mathrm{i}}{L}\right)\nonumber\\
&&\qquad\qquad-\left(\frac{\omega M_0}{R_0}\right)^2\left(\frac{R_0}{R_\delta}+\frac{1}{2}\frac{L_0}{L}\right)
\end{eqnarray}
where $\omega_0$ and $Q$ are the intrinsic SRR values in the limit that $M_\mathrm{i}$ and $M_0$ go to zero.

Note that, while both $\omega_\mathrm{eff}$ and $Q_\mathrm{eff}$ deviate from their zero-coupling values as the square of the mutual inductances $M_\mathrm{i}$ and $M_0$, the coefficients in front of these factors differ.  For a typical SRR, one has \mbox{$L_\mathrm{i,0}/L\ll 1$} and \mbox{$R_\mathrm{i,0}/R_\delta\gg 1$}.  As a result, while the resonator's $Q$ can decrease significantly as the coupling strength is increased, the relative change in its resonance frequency is expected to be much less.

\subsection{Measured Coupling Dependence of the Toroidal SRR}

\begin{figure}[t]
\includegraphics[keepaspectratio, width=\columnwidth]{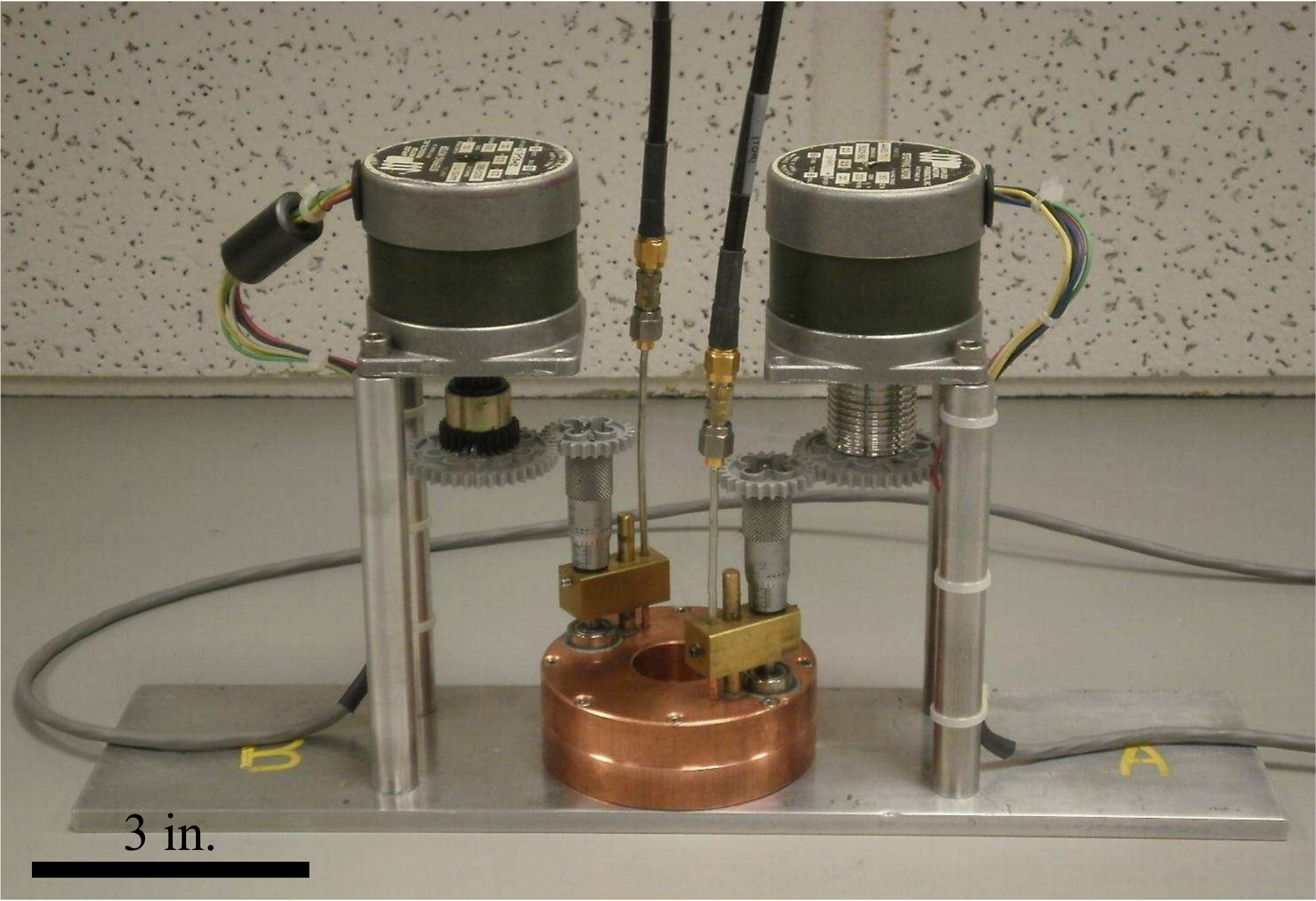}
\caption{\label{fig:step}Toroidal SRR with the coupling loop positions set using micrometers with 1~in.\ of travel. Adjustments to the coupling loop positions were automated using stepper motors and a LabVIEW program. The black line in the bottom-left corner of the figure represents a length of 3~inches.}
\end{figure}

\begin{figure*}
\begin{tabular}{lclc}
(a) & \includegraphics[keepaspectratio, width=0.94\columnwidth]{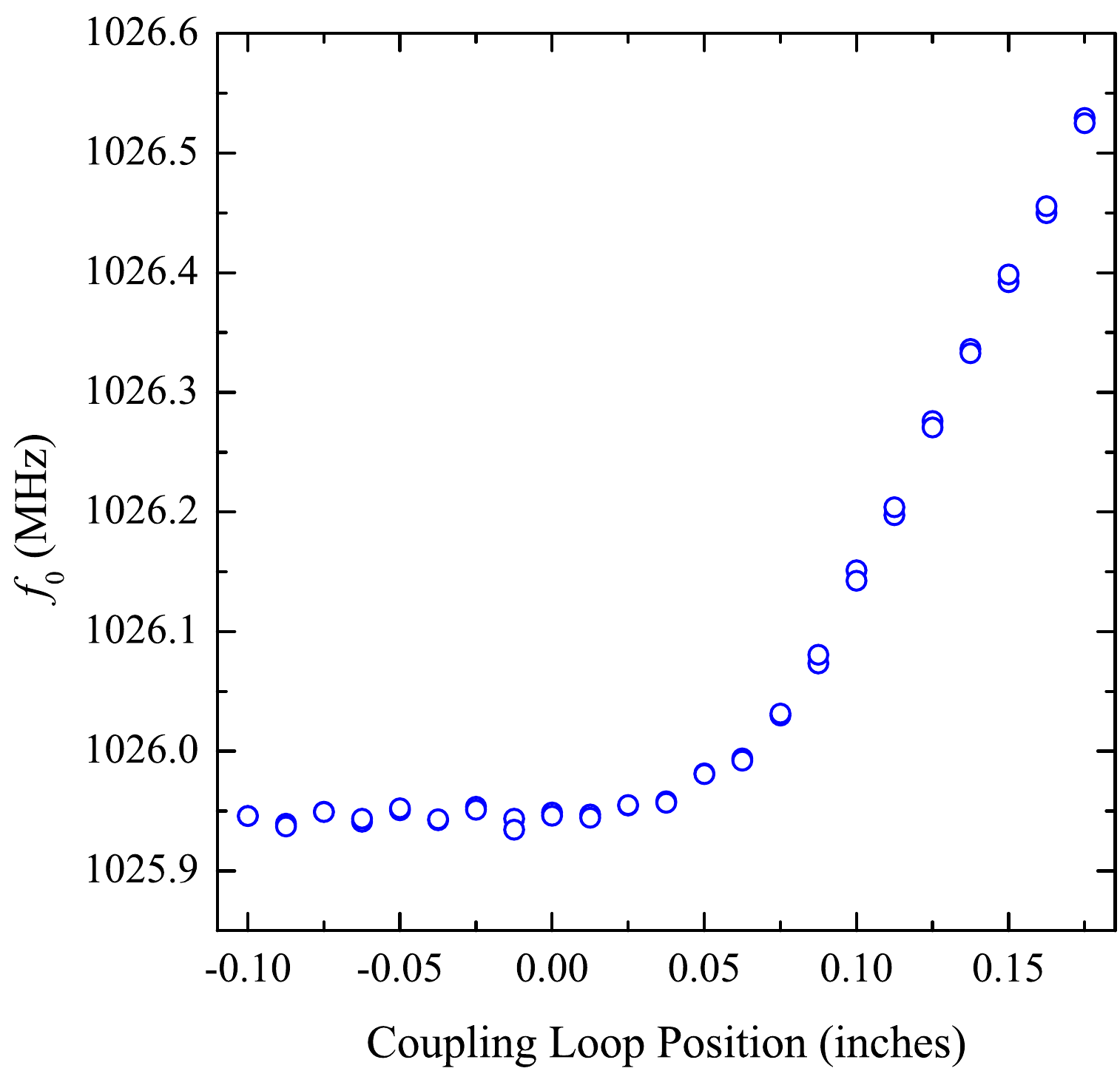} &
(b) & \includegraphics[keepaspectratio, width=0.94\columnwidth]{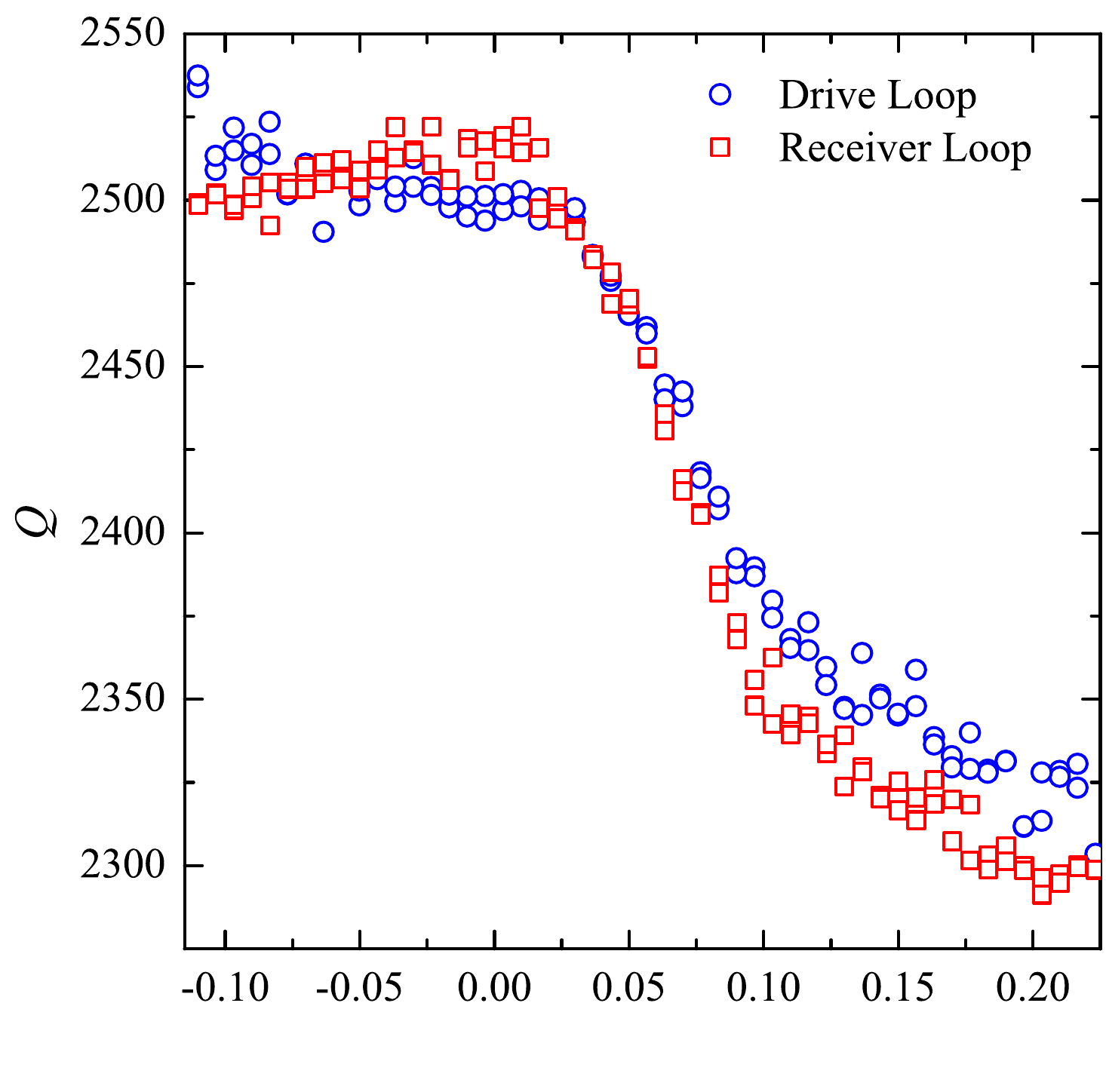}
\end{tabular}
\caption{\label{fig:scan}(a) Toroidal SRR resonance frequency measured as a function of the position of the drive coupling loop.  (b) Measured toroidal SRR quality factor versus the positions of the drive (blue circles) and receiver (red squares) coupling loops.}
\end{figure*}

Translation stages that allowed the torodial SRR coupling loop positions to be accurately and reproducibly set were built. As shown in Fig.~\ref{fig:step}, the coupling loops were anchored to stages that can be vertically translated using micrometers with a resolution of 0.001~in.\ (25~$\mu$m).  The process of adjusting the coupling loop positions was automated by epoxying plastic gears to the micrometers and using a pair of stepper motors controlled by a LabVIEW program and controller circuit designed in-house.    

Figure~\ref{fig:scan} shows how $f_0$ and $Q$ varied during scans of the coupling loop positions.  For these measurements, the input signal was supplied using the Rohde \& Schwarz SMY02 signal generator and the signal coupled out of the SRR was detected using the Anritsu MS610C2 spectrum analyzer.  In the figure, a position of zero corresponds to the point at which the entire coupling loop has just been extracted from the bore of the SRR.  Negative positions indicate that the coupling has been pulled further into to the access hole drilled through the top of the resonator.  For positive positions, the coupling loop is protruding into the bore of the resonator.  At positions greater than approximately 0.080~in.\ (2.0~mm), the coupling loop is entirely within the bore of the resonator (maximum coupling).    

Figure~\ref{fig:scan}(a) shows that, as the drive coupling loop is moved towards the bore of the resonator, with the receiver loop position fixed at zero, $f_0$ is initially insensitive to changes in its position.  However, for positions greater than $0.08$~in., $f_0$ rises sharply as the loop is pushed further into the bore.  As was anticipated in the previous section, $f_0$ responds very weakly to changes in the coupling strength because \mbox{$L_\mathrm{i}/L_\mathrm{t}\ll 1$}.  The relatively large increase in $f_0$ observed once maximum coupling is reached is due a change in the resonator's inductance $L_\mathrm{t}$.  As the coupling loop is pushed further into the SRR, the outer conductor of the coaxial cable enters the bore of the resonator.  Due to the skin effect, magnetic flux is excluded from the volume of the bore occupied by the coaxial cable thereby lowering the effective SRR inductance and increasing its resonance frequency.  The linear increase in $f_0$ is expected because, for positions greater than $0.080$~in., the excluded volume increases linearly with position.

Figure~\ref{fig:scan}(b) examines how the $Q$ of the toroidal SRR varies with coupling loop position.  Initially, with the drive (receiver) coupling loop fully extracted deep within the hole through the top half of the SRR, changes to its position result in very little change to the mutual inductance $M_\mathrm{i}$ ($M_0$) and, hence, no noticeable change to $Q$.  As the position is increased from 0 to 0.08~in., the coupling loop goes from being completely outside to being completely within the bore of the resonator. Associated with this change in position is a large increase in the coupling between the SRR and external circuits and, hence, a decrease in the $Q$ of the resonator.  Increasing the position of the coupling loop beyond 0.08~in.\ does not result in stronger coupling because, at this point, the coupling loop is entirely within the bore of the resonator and the magnetic flux through the loop is approximately constant.  Although the coupling remains constant for positions greater than 0.08~in., the $Q$ of the resonator continues to decrease slowly as the position is further increased.  This decrease in $Q$ is due to power losses associated with eddy currents induced on the outer conductor of the coaxial cable as it enters the bore of the SRR.

\begin{figure*}
\begin{tabular}{lclc}
(a) & \includegraphics[keepaspectratio, width=.97\columnwidth]{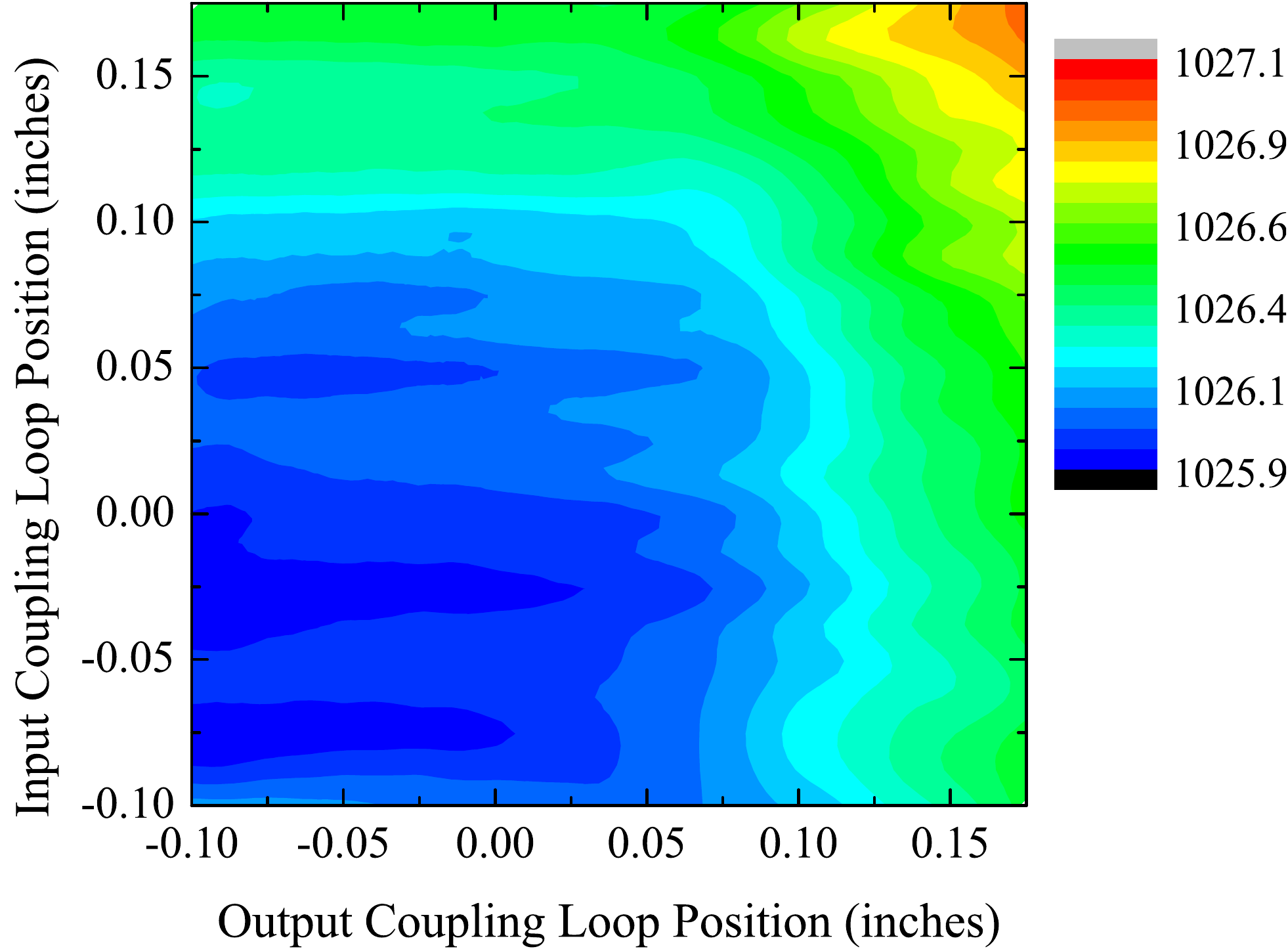} & (b) & \includegraphics[keepaspectratio, width=.97\columnwidth]{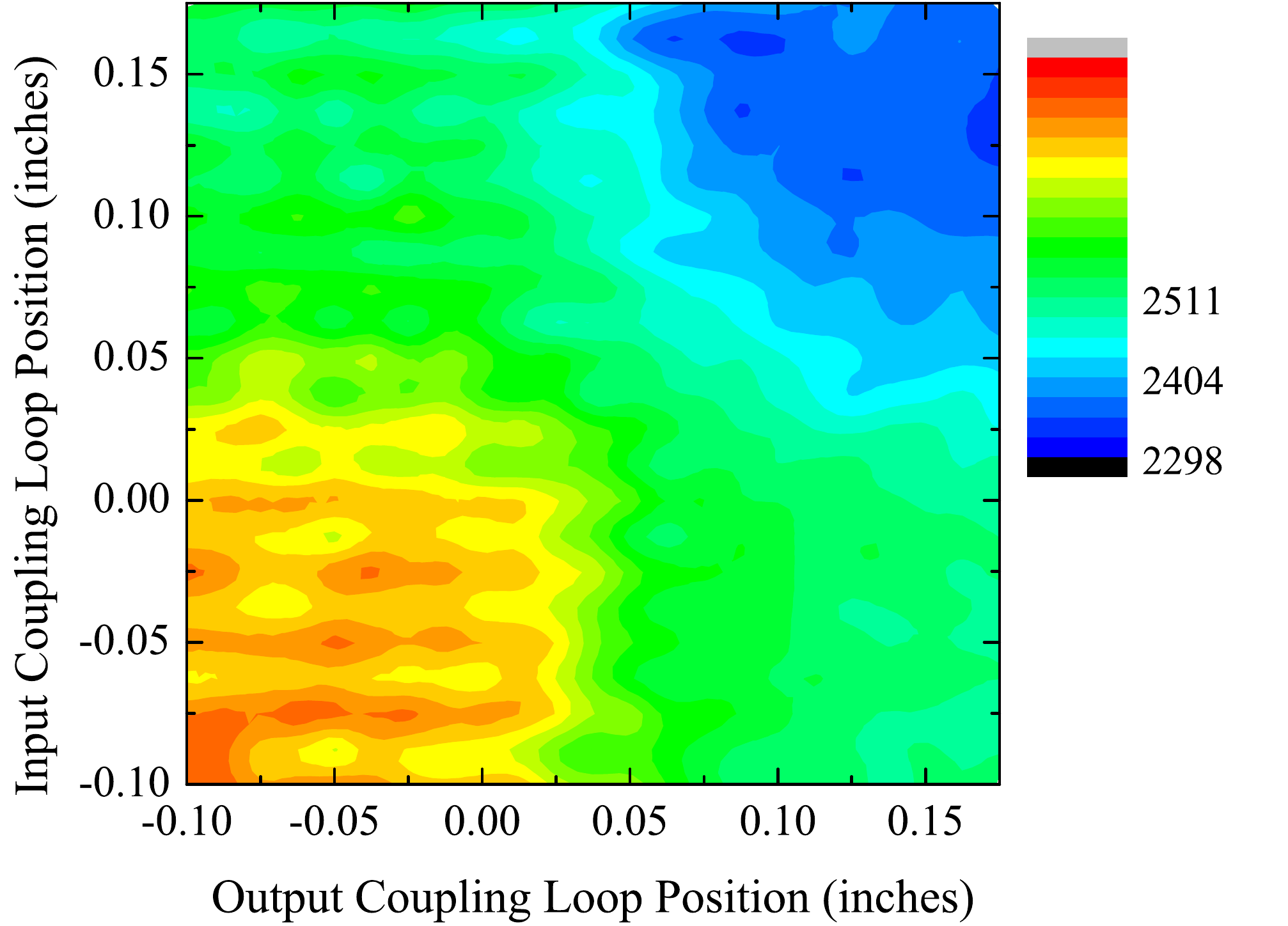}
\end{tabular}
\caption{\label{fig:contour}Contour plots of (a) the resonant frequency in MHz and (b) the $Q$ of the toroidal SRR as a function of the coupling loop positions.}
\end{figure*}

Finally, contour plots showing how $f_0$ and $Q$ of the toroidal SRR vary with the positions of the coupling loops are presented in Fig.~\ref{fig:contour}.  The resonance frequency and quality factor of the resonator were measured for more than 1400 unique sets of coupling loop positions.  Figure~\ref{fig:contour}(a) shows that the resonance frequency remains approximately constant provided that the outer conductors of the coupling loops do not enter the bore of the resonator (positions less than 0.08~in.).  When the outer conductors are within the bore of the resonator, the increase in $f_0$ is determined by the fraction of the bore's volume that is occupied by the coaxial cables.  A pair of 0.085~in.\ diameter coaxial cables that extend 0.095~in.\ (2.41~mm) into the bore of the resonator occupy approximately 0.087\% of the SRR's active volume.  This rough estimate agrees reasonably well with the observed 0.097\% increase in $f_0$.

Figure~\ref{fig:contour}(b) shows that the $Q$ of the SRR is at its maximum and approximately constant when the coupling loops are both outside the bore of the resonator (bottom-left red and yellow patch).  The $Q$ drops smoothly has the coupling loop positions are increased due to an enhanced mutual inductance between the SRR and external circuits used to generate and detect signals.  After both loops are fully within the bore of the resonator, the $Q$ continues to drop, but much more slowly, due to eddy currents induced in the outer conductors of the coaxial cables (top-right blue patch).  Note that both Figs.~\ref{fig:contour}(a) and (b) are symmetric about a diagonal line with slope one.  This symmetry indicates that the drive and receiver loops affect the SRR in the same way which is expected when one has nearly identical coupling loops and \mbox{$R_\mathrm{i}=R_0=50~\Omega$}.

\subsection{Capacitive Coupling}
All of the data presented in this work were obtained by inductively coupling the SRR to external circuits.  However, it is also possible to capacitively couple electric flux in the gap of the SRR to external circuits.\cite{Hardy:1981}  Capacitive coupling can be achieved using a coaxial cable with an open-circuit termination.  At the open end of the coaxial cable, electric field lines radiate out into free space.\cite{WhitAthey:1982, Bobowski:2012}  If the open end of the cable is placed in the vicinity of the gap of the SRR, electric flux can be coupled into and out of the resonator.

An equivalent circuit analogous to the one shown in Fig.~\ref{fig:couplingCircuit} can be developed for capacitive coupling.  Analysis of the circuit model leads to expressions that determine how the SRR's resonant frequency and quality factor depend on the coupling strength.  An increase in the coupling strength lowers both the resonant frequency and $Q$ of a capacitively coupled SRR.\cite{Bobowski:2016}  We have used an open-ended coaxial cable to capacitively couple to the toroidal SRR.  Access holes drilled into the top ring of the toroidal SRR allowed the open end of a coaxial cable to be placed near or within the gap of the resonator.  We measured quality factors that were similar to those obtained using inductive coupling.  We also confirmed that both $f_0$ and $Q$ decreased as the coupling strength was increased.\cite{Bobowski:2016}

\section{Summary}\label{sec:summary}

A practical toroidal SRR was designed, built, and characterized.  The novel toroidal geometry strongly confines the magnetic flux within the bore of the resonator and thus avoids the radiative power losses associated with conventional cylindrical SRRs.  Compact toroidal SRRs can be designed to have resonance frequencies from tens of megahertz up to several gigahertz.  Compared to traditional cylindrical SRRs, toroidal SRRs maintain high quality factors and very stable resonance frequencies without requiring additional EM shielding.

In this paper, the intrinsic capacitance, inductance, and resistance of the toroidal SRR were calculated and used to estimate the expected resonance frequency and quality factor of the resonator.  The magnitude and phase of the resonator's frequency response was measured using a vector network analyzer and its transient response to an rf pulse was measured using a high-frequency oscilloscope.  The experimentally determined resonance frequency $f_0=1.03$~GHz was very close to the design value of 1.06~GHz and the measured $Q$, under weak-coupling conditions, was only 15\% below the calculated value.  Both the frequency and transient responses can be used to fully characterize the SRR.

Next, the effect of inductively coupling the SRR to external excitation and detection circuits was investigated.  In the limit that the coupling loop inductances are small, it was shown that $f_0$ and $Q$ deviate from their zero-coupling values as the square of the input and output mutual inductances $M_\mathrm{i}$ and $M_0$.  However, the lowest order corrections to $f_0$ go as \mbox{$L_\mathrm{i}/L$, $L_0/L\ll 1$} whereas the corrections to $Q$ go as \mbox{$R_\mathrm{i}/R_\delta$, $R_0/R_\delta\gg 1$}.  

Experimental measurements confirmed that $f_0$ did not vary with the strength of the coupling.  However, $f_0$ was observed to rise sharply once the outer conductors of the coaxial cables used to make the coupling loops entered the bore of the resonator.  The presence of the coaxial cables reduces the effective volume of the resonator's bore and therefore its inductance.  As predicted, the $Q$ of the SRR was observed to decrease as the coupling strength, to either the drive or detector circuit, was increased.      

Here, we comment briefly on the use of the SRRs to measure electromagnetic material properties.  The capacitance of the SRR can be modified by filling the gap of the resonator with a dielectric material with complex relative permittivity $\varepsilon_\mathrm{r}$ and the inductance can be altered by filling the bore of the resonator with a ferromagnetic material or suspension with complex relative permeability $\mu_\mathrm{r}$. Reference~\onlinecite{Bobowski:2015} describes in detail how the real and imaginary components of $\varepsilon_\mathrm{r}$ and $\mu_\mathrm{r}$ can be obtained from successive measurements of $f_0$ and $Q$ with the resonator in-air and then with its gap or bore filled with the material of interest.  In Ref.~\onlinecite{Bobowski:2013}, a cylindrical SRR was used to measure the dielectric constant and conductivity of water with various concentrations of dissolved salt.   Cylindrical SRRs have also been used to characterize the magnetic penetration depth and surface resistance of high-temperature superconductors.\cite{Bonn:1991, Hardy:1993, Bobowski:2010}  We have successfully used the toroidal SRR to measure the complex permittivity of methanol at 185~MHz, the dielectric constant of liquid nitrogen from 63 to 78~K, and the conductivity of copper from 77 to 300~K.  The details of these measurements will be the topic of a future publication.   

We are currently developing a low-frequency electron spin resonance (ESR) apparatus using the toroidal SRR described in this paper.  The toroidal SRR geometry allows one to design a high-$Q$ and compact resonator with a high filling factor that operates near 1~GHz.  An ESR apparatus that operates at low frequencies can have both practical and scientific advantages. The static magnetic field required for ESR measurements is directly proportional to the rf frequency used.  The static magnetic field required for ESR experiments at 1~GHz would be approximately 30--40~mT.  Fields of this magnitude can be generated easily using inexpensive electromagnets built in-house.  Furthermore, some ESR spectra have field-dependent lineshapes such that valuable information can be obtained by studying these spectra at a number of different static magnetic fields and, therefore, frequencies.\cite{Momo:1983}

Finally, we note that miniaturized toroidal SRRs could find applications in the very active field of metamaterials.  Meta-atoms based on toroidal SRRs would have very high intrinsic quality factors.  Coupling between adjacent toroidal SRRs could be achieved via optical coupling as in Ref.~\onlinecite{Slobozhanyuk:2014}.  Alternatively, if the the gap of the toroidal SRR was located at the outer diameter resonator, it may be possible to use the fringing electric fields to couple adjacent toroidal SRRs.

\begin{acknowledgments}
We wish to acknowledge the support of Thomas Johnson and Jonathan Holzman who generously provided access to the Agilent N5241A VNA and Tektronix DP070804B oscilloscope, respectively.  We also acknowledge the support provided by Durwin Bossy and the UBC Okanagan machine shop.
\end{acknowledgments}

\appendix*

\section{Toroidal SRR Resistance}\label{Append:Rt}

\begin{figure}[t]
\includegraphics[keepaspectratio, width=\columnwidth]{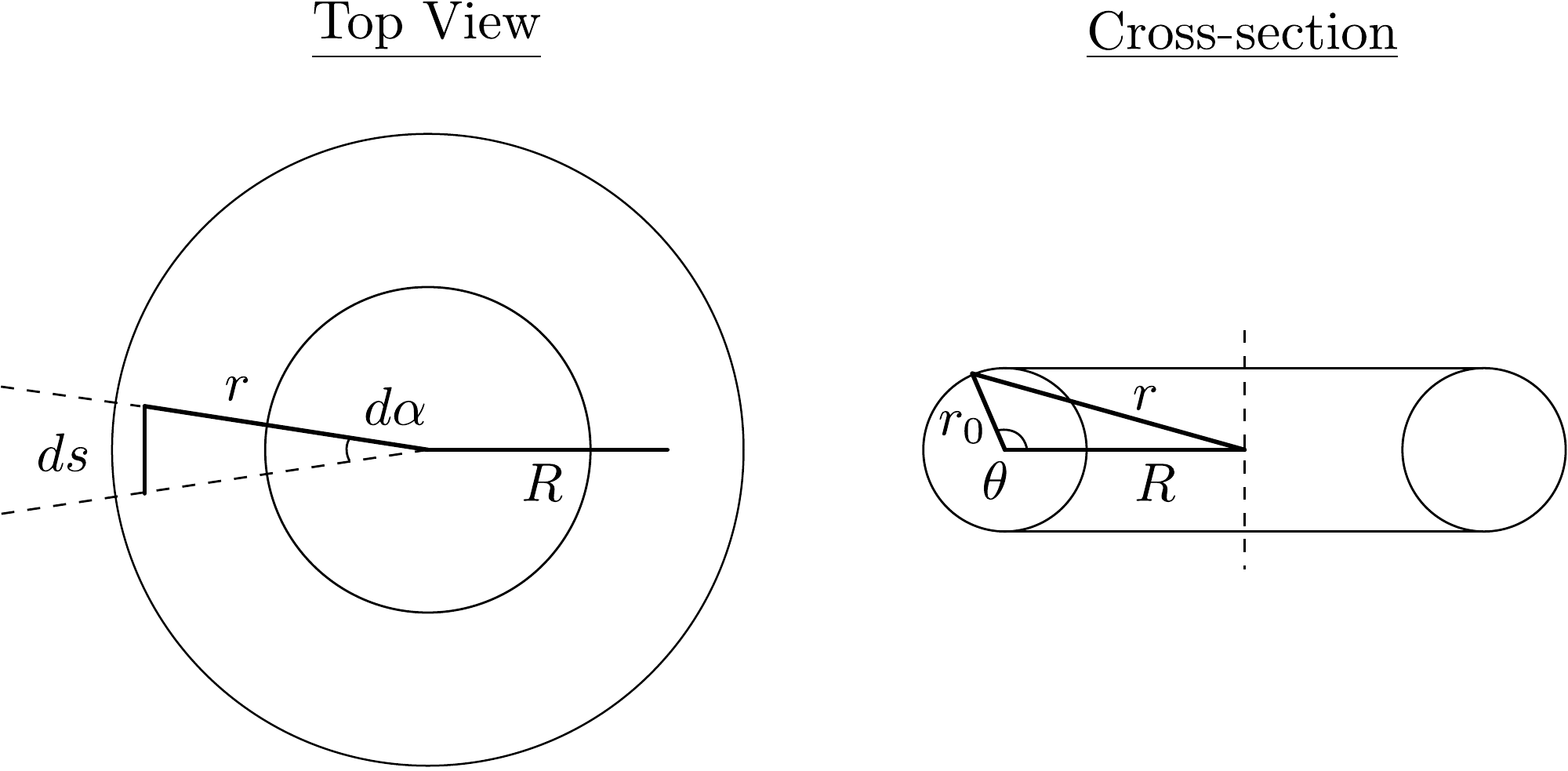}
\caption{\label{fig:A1}Geometry of the toroidal SRR bore used to calculate the effective resistance of the resonator.  The toroid has radius $R$ and the radius of the bore is $r_0$.}
\end{figure}

In this appendix the effective resistance of the toroidal SRR is calculated.  Here, we are concerned with the intrinsic resistance of the SRR due to the resistivity $\rho$ and the skin depth $\delta$ of the conductor used to make the resonator.  First, the resistance of a narrow strip of circular cross-section and thickness $\delta$ is calculated.  The net resistance is then determined from a parallel combination of many strips used to form the toroid.  Figure~\ref{fig:A1} shows the geometry of the problem.

The width of a strip of the toroid of angular size $d\alpha$ is \mbox{$ds\approx r(\theta) d\alpha$} which varies with angle $\theta$.  As a result, for currents running along the length of the strip, the effective resistance is given by
\begin{equation}
dR_\mathrm{t}\approx \frac{\rho r_0}{\delta R\,d\alpha}\int^{2\pi}_0\frac{d\theta}{\sqrt{1+\left(r_0/R\right)^2-2\left(r_0/R\right)\cos\theta}}.
\end{equation} 
where $r(\theta)$ has been expressed in terms of $R$, $r_0$, and $\theta$ using the cosine law.
The net resistance $R_\mathrm{t}$ of the resonator is determined from the parallel combination of strips used to form the complete toroid \mbox{$R_\mathrm{t}^{-1}=\displaystyle\int_{\alpha=0}^{2\pi} dR_\mathrm{t}^{-1}$} such that:
\begin{equation}
R_\mathrm{t}=\frac{\rho\, r_0}{2\pi R\,\delta}\bigintsss_0^{2\pi}\frac{d\theta}{\sqrt{1+\left(r_0/R\right)^2-2\left(r_0/R\right)\cos\theta}}\label{eq:A}
\end{equation}
which can be evaluated numerically for a given value of $r_0/R$.

\nocite{*}
\bibliography{Bobowski-RSI-ToroidalSRR-20160103}

\end{document}